\documentclass[12pt]{article}
\usepackage{amsmath}
\usepackage{amsfonts}
\usepackage{graphicx}
\usepackage{mathrsfs}

\begin{document}

%
\def\papertitlepage{\baselineskip 3.5ex \thispagestyle{empty}}
\def\preprinumber#1#2{\hfill \begin{minipage}{4.2cm}  #1
                 \par\noindent #2 \end{minipage}}
\renewcommand{\thefootnote}{\fnsymbol{footnote}}
\newcommand{\beq}{\begin{equation}}
\newcommand{\eeq}{\end{equation}}
\newcommand{\beqa}{\begin{eqnarray}}
\newcommand{\eeqa}{\end{eqnarray}}
\catcode`\@=11
\@addtoreset{equation}{section}
\catcode`@=12
\relax
%
%
%
\papertitlepage
\setcounter{page}{0}
\preprinumber{KEK-TH-1414}{}
\baselineskip 0.8cm
\vspace*{2.0cm}
\begin{center}
{\large\bf Note on Anomaly Cancellation  \\
on $SO(32)$ heterotic 5-brane}
\end{center}
\vskip 4ex
\baselineskip 1.0cm
\begin{center}
           {Harunobu Imazato${}^{\sharp}$\footnote{imaza@post.kek.jp}, Shun'ya Mizoguchi${}^{\sharp~\!\!\flat}$\footnote{mizoguch@post.kek.jp}
           and Masaya Yata${}^{\sharp}$\footnote{yata@post.kek.jp}} 
\\
\vskip 1em
     ${}^\sharp$  {\it 
Department of Particle and Nuclear Physics}\\
 \vskip -2ex {\it The Graduate University for Advanced Studies}\\
 \vskip -2ex {\it Tsukuba, Ibaraki 305-0801, Japan} \\
\vskip 1em
      ${}^\flat$ {\it Theory Center}\\
 \vskip -2ex {\it High Energy Accelerator Research Organization (KEK)} \\
       \vskip -2ex {\it Tsukuba, Ibaraki 305-0801, Japan} \\
\end{center}
\vskip 5ex
%
\baselineskip=3.5ex
\begin{center} {\bf Abstract} \end{center}

\vskip 2ex

We show that the gauge, gravitational (tangent-bundle) and their mixed 
anomalies arising from the localized modes near a 5-brane 
in the $SO(32)$ heterotic string theory cancel with the anomaly 
inflow from the bulk with the use of the Green-Schwarz mechanism on the brane, 
similarly to the $E_8\times E_8$ 5-brane case. 
We also compare our result with Mourad's analysis 
performed in the small-instanton limit.

\vspace*{\fill}
\noindent
October 2010
\newpage
\renewcommand{\thefootnote}{\arabic{footnote}}
\setcounter{footnote}{0}
\setcounter{section}{0}
\baselineskip = 0.6cm
\pagestyle{plain}

\indent One of the most amazing aspects of string theory is the miraculous mechanism 
of anomaly cancellation. In the ten-dimensional bulk, the anomalies of $N=1$ 
superstrings are successfully cancelled by the use of the well-known Green-Schwarz 
mechanism \cite{GSmechanism}, in which the two-form $B$ field is assumed to 
change with respect to the gauge and local Lorentz transformations. 
Anomalies of chiral matter fields supported on some branes are also known to 
cancel with inflow contributions from the bulk \cite{anomaly_inflow}. 
In this letter, we focus on the gauge, gravitational (tangent-bundle) 
and their mixed anomalies 
arising from the localized modes near a 5-brane in the $SO(32)$ heterotic 
string theory. We show that their anomalies also cancel with the anomaly 
inflow from the bulk with the use of the Green-Schwarz mechanism on the brane, 
similarly \cite{KM} to the case of the $E_8\times E_8$ 5-brane. 
Although the argument that the anomalies on a heterotic 5-brane should cancel 
with an anomaly inflow is an old one \cite{BlumHarvey}, the arithmetic we show below 
is new and different from \cite{BlumHarvey}, as, for instance, we do not consider 
any ``current at infinity".
We also compare our result with Mourad's analysis \cite{Mourad} 
in which the small-instanton limit was considered.
Anomaly cancellation on heterotic 5-branes in the K3 compactification was discussed in \cite{Honecker}.

\indent
Let us start with the symmetric 5-brane solution 
\cite{CHS,SJRay} in the $SO(32)$ heterotic string theory.
It has been known for some time that the moduli of this solution consists of $D=6$, 
${\cal N}=1$ 30 hypermultiplets \cite{CHS2}. The bosonic moduli are 
four Nambu-Goldstone modes associated with the spontaneously broken translational 
invariance, one scale modulus and 115 moduli coming from the arbitrariness of 
the choice of $SU(2)$ subgroup of $SO(32)$ in which the spin connection is 
embedded. The number of 115 can be easily counted by the decomposition of $SO(32)$ 
in terms of $SO(28)\times SU(2) \times SU(2)$ as follows: 
\beqa
{\bf 496}&=&({\bf 378},{\bf 1},{\bf 1})\otimes
({\bf 1},{\bf 3},{\bf 1})\otimes
({\bf 1},{\bf 1},{\bf 3})\otimes
({\bf 28},{\bf 2},{\bf 2}).
\eeqa
Suppose that the $SU(2)$ spin connection is embedded into the last $SU(2)$ subgroup.
Then the centralizer $SO(28)\times SU(2)$ remains as the unbroken gauge group, while 
the rest of $3+28\times 2\times 2=115$ generators give rise to deformations, being moduli
of this solution.

\indent
Thus, 28 of the 30 hypermultiplets, which contain 56 symplectic Majorana-Weyl spinors,  
transform as $({\bf 28},{\bf 2})$ with respect to the unbroken $SO(28)\times SU(2)$ 
gauge symmetry, while the remaining two are gauge singlets. 
Anomaly polynomials 
for the chiral fermions belonging to these hypermultiples are:
\beqa
I_8^{({\bf 28},{\bf 2})}&=&\left.\frac12 \hat{A}(T\Sigma)\cdot
\mbox{tr}_{({\bf 28},{\bf 2})}e^{iF}\right|_8\nonumber\\
&=&\frac{28}{5760}(-4 p_2 + 7 p_1^2)
+\frac1{96}p_1\mbox{tr}_{({\bf 28},{\bf 2})}F^2
+\frac1{48}\mbox{tr}_{({\bf 28},{\bf 2})}F^4,
\label{I^bifundamental}\\
I_8^{singlet}&=&\left.\frac12 \hat{A}(T\Sigma)\right|_8\times 4
\nonumber\\
&=&\frac{2}{5760}(-4 p_2 + 7 p_1^2),
\label{I^singlet}
\eeqa
where $\hat{A}(T\Sigma)$ is the Dirac genus of the tangent bundle of the 5-brane 
and $F$ is the 2-form for the $SO(28)\times SU(2)$ gauge field strength.
Similarly to \cite{KM}, we have ignored the normal bundle anomalies in 
(\ref{I^bifundamental}) and (\ref{I^singlet}). 
The total anomaly $I_6^1$ is obtained by the well-known descent relations:
\beqa
I_8&=&I_8^{({\bf 28},{\bf 2})}+I_8^{singlet},\\
I_8&=&dI_7,\\
\delta I_7&=&dI_6.
\eeqa

\indent
On the other hand, the bulk supergravity action contains the Green-Schwarz 
counterterm proportional to $BX_8$ with
\beqa
X_8&=&\frac1{24}\left(\frac18 \mbox{tr}R^4
+\frac1{32} (\mbox{tr}R^2)^2
-\frac1{240}\mbox{tr}R^2\mbox{Tr}F_{SO(32)}^2
+\frac1{24}\mbox{Tr}F_{SO(32)}^4
-\frac1{7200}(\mbox{Tr}F_{SO(32)}^2)^2
\right),\nonumber\\
\eeqa
where $F_{SO(32)}$ is the 2-form for the $SO(32)$ gauge field strength 
and $\mbox{Tr}$ is the trace in the adjoint ${\bf 496}$ representation. 
$R$ is the curvature 2-form for the ten-dimensional bulk tangent bundle 
$Q$, which is decomposed, in the presence of the 5-brane, into a direct sum 
of the tangent bundle of the brane, $T\Sigma$, and the normal bundle $N$ of it.  
A Pontryagin class of $Q$ can be expressed as a polynomial of Pontryagin 
classes of $T\Sigma$ and $N$. However, since we have taken only the
tangent bundle anomalies into account in (\ref{I^bifundamental}) and (\ref{I^singlet}), 
we may identify the curvature 2-form $R$  for the total bundle $Q$ to be 
the curvature 2-form for the tangent bundle $T\Sigma$ of the brane, and 
examine the cancellation of the tangent 
bundle anomalies, as well as the gauge anomalies, and the mixed ones 
for the gauge and tangent bundles. In the end of this note, we also 
comment on the normal bundle anomaly cancellation in the present setting.
With these remarks we write $X_8$ in terms of traces of the subgroup
$SO(28)\times SU(2)$:
\beqa
X_8&=&\frac1{192}\left(\rule{0ex}{3ex}
3p_1^2 - 4 p_2 + 2 p_1 
(\mbox{tr}_{\bf 28}F_{SO(28)}^2 +2  \mbox{tr}_{\bf 2}F_{SU(2)}^2)
+8( \mbox{tr}_{\bf 28}F_{SO(28)}^4 +2 \mbox{tr}_{\bf 2}F_{SU(2)}^4)
\right),
\nonumber\\
\label{X8}
\eeqa
where $p_1$ and $p_2$ are now understood as the Pontryagin classes 
for $T\Sigma$.

\indent
The gauge and local Lorentz variations of the B field in the $BX_8\sim - dB X_7$ 
term precisely cancel the ten-dimensional bulk anomalies in the $SO(32)$ 
string theory; this is the Green-Schwarz mechanism. On the other hand, 
if there is no brane, the variations of $X_7$ vanishes because $d^2 B=0$. 
However, since the 5-brane is a magnetic source for the $B$ field, the variations of 
$X_7$ give rise to, in the presence of the 5-brane, 
$\delta$-function-like contributions on the 5-brane known as anomaly inflows. 
Therefore, the total anomalies are described by the invariant polynomial
$I_8^{({\bf 28},{\bf 2})}+I_8^{singlet} - X_8$, which turns out, using 
(\ref{I^bifundamental}), (\ref{I^singlet}) and ({\ref{X8}}), to factorize as
\beqa
I_8^{({\bf 28},{\bf 2})}+I_8^{singlet} - X_8 &=& 
-\frac1{96}\left(\rule{0ex}{3ex}\mbox{tr}R^2 - \mbox{tr}_{\bf 32}F_{SO(32)}^2
\right)
\left(\rule{0ex}{3ex}p_1 
+ 12  \mbox{tr}_{\bf 2}F_{SU(2)}^2
\right). \label{heteroticSO(32)anomaly}
\eeqa 
Here we have reexpressed $\mbox{tr}_{\bf 28}F_{SO(28)}^2 
+2  \mbox{tr}_{\bf 2}F_{SU(2)}^2$ as the $SO(32)$ fundamental trace in the 
first parentheses. The first factor is precisely the combination that appears 
in the anomalous Bianchi identity of the $H$ field, and therefore the sum of 
anomalies can be cancelled by introducing a local counterterm on the 5-brane, 
similarly to the cases of the type I \cite{Mourad} and $E_8 \times E_8$ \cite{KM} 
5-branes.

\indent
Let us compare the arithmetic we presented above with the known 
mechanism of anomaly cancellation on the $SO(32)$ 5-brane 
in the small-instanton limit \cite{Mourad}. If the instanton size of the heterotic 
5-brane tends to zero, the theory becomes strongly coupled and the supergravity 
analysis loses its validity. It was proposed that there 
would then be an enhanced $Sp(k)$ gauge symmetry \cite{Witten} on $k$ parallel 
5-branes in this limit of the $SO(32)$ theory, in addition to the full $SO(32)$ 
gauge symmetry. This argument was supported by the 
proof of anomaly cancellation in the S-dual type I brane system \cite{Mourad}:
The dual type I D5-branes have three kinds of zeromode hypermultiplets, called 
$\theta$, $\lambda$ and $\psi$ in \cite{Mourad}; they transform as 
$({\bf 4_+},{\bf 2_+},{\bf 1},{\bf 1})$, 
$({\bf 4_-},{\bf 2_-},{\bf 1},{\bf 3})$ and 
$({\bf 4_+},{\bf 1},{\bf 32},{\bf 2})$,
respectively, under the actions of $SO(5,1)\times SO(4) \times SO(32) \times SU(2)$,
where 
the first two factors are the ten-dimensional Lorentz group, while the last 
$SU(2)=Sp(1)$ is the enhanced gauge symmetry. The subscripts $\pm$ denote
the chiralities of the spinors.

Let $I_8^\theta$, $I_8^\lambda$ and $I_8^\psi$ be the anomaly polynomials of 
$\theta$, $\lambda$ and $\psi$, respectively. If $k=1$, the total anomaly turns 
out to be \cite{Mourad}
\begin{eqnarray}
I_8^\theta + I_8^\lambda + I_8^\psi-X_{8}
=-\frac{1}{96}\left(\rule{0ex}{3ex}
\mbox{tr}R^{2}-\mbox{tr}_{\bf 32} F_{SO(32)}^{2}
+4\chi(N)\right)
\left(\rule{0ex}{3ex}p_{1}(Q)+ 12\mbox{tr}_{\bf 2} F^{2}_{SU(2)}-2p_{1}(N)\right),
\label{D5anomaly}
\end{eqnarray}
where $\chi(N)$ is Euler class of the normal bundle. 

If all the terms depending on the normal-bundle connections are 
ignored in (\ref{D5anomaly}), then $p_1(Q)$ is replaced with $p_1(T\Sigma)$, and 
(\ref{D5anomaly}) looks superficially the same as (\ref{heteroticSO(32)anomaly}).  
There are, however, a number of significant differences between 
our result and Mourad's analysis as follows:
\begin{itemize}
\item[(i)]In (\ref{heteroticSO(32)anomaly}), {$F_{SU(2)}$ is the field strength of 
the unbroken $SU(2)$ {\em subgroup} of $SO(32)$, whereas
in (\ref{D5anomaly}) is that of the {\em enhanced} $SU(2)$ gauge group
which is independent of the bulk $SO(32)$ gauge symmetry. 
}
\item[(ii)]{Although one could decompose $SO(32)$ representations into 
those of the subgroup $SO(28)\times SU(2)$ in (\ref{D5anomaly}), 
the supermultiplets turn out to transform quite differently from those 
in our ``broken" case. For instance, we have an $SO(28)\times SU(2)$ 
bifundamental, while there arise no such representations in the small-instanton 
case.}
\item[(iii)]{Mourad's proof of cancellation extends to $k(\geq 2)$ $D5$-branes, 
while it is not obvious to generalize our argument to the case of many heterotic 5-branes.
}
\item[(iv)]{
It is also difficult to include the normal-bundle contributions 
in our case; a naive inclusion of them does not lead to the desired factorized 
form.  Since it is known that the mechanisms of normal bundle anomaly cancellation 
on M5-branes requires a complicated setting \cite{FHMM}, we might also need 
to consider in heterotic string theories such a modification of the solution 
without small instantons.
}
\end{itemize}

Since the heterotic/type I duality is a strong-weak duality, there is no guarantee that 
how anomalies cancel in one theory will be the same in the other theory.  
Therefore, we conclude that the superficial similarity between 
(\ref{heteroticSO(32)anomaly}) and
(\ref{D5anomaly}) will be an accident for $k=1$.

\section*{Acknowledgments}

\hskip 0.2 in
We thank T.~Kimura for helpful discussions. 
We also thank Y.~Imamura, H.~Itoyama, Y.~Kitazawa, K.~Murakami, 
T.~Sasaki, I.~Tsutsui and Y.~Yasui for discussions and comments.
The work of S.~M. is supported by 
Grant-in-Aid
for Scientific Research (C) \#20540287-H20 and 
(A) \#22244430-0007
from
The Ministry of Education, Culture, Sports, Science
and Technology of Japan.
%

\end{document}